\documentclass[preprint,12pt]{elsarticle}
\usepackage{graphicx,amssymb}
\journal{(Physica D)}
\def\be{\begin{equation}}
\def\ee{\end{equation}}
\def\bea{\begin{eqnarray}}
\def\eea{\end{eqnarray}}

\def\e{\mathrm{e}}

\def\pd{\partial}

\def \l{{\cal L}}

\begin{document}

\begin{frontmatter}

\title{Dynamics of DNA breathing in the Peyrard-Bishop model with damping and
external force}

\author[bppt]{A. Sulaiman\footnote{Email : albertus.sulaiman@bppt.go.id}} 
\author[itb,icmp]{F.P Zen\footnote{Email : fpzen@fisika.itb.ac.id}} 
\author[ipb,icmp]{H. Alatas\footnote{Email : alatas@ipb.ac.id}} 
\author[gftk,ui]{L.T. Handoko\footnote{Email : handoko@teori.fisika.lipi.go.id,
laksana.tri.handoko@lipi.go.id}}
\address[bppt]{Badan Pengkajian dan Penerapan Teknologi, BPPT Bld. II (19$^{\rm
th}$ floor), Jl. M.H. Thamrin 8, Jakarta 10340, Indonesia}
\address[itb]{Theoretical High Energy Physics and Instrumentation (THEPI),
Department of Physics, Institut Teknologi Bandung, Jl. Ganesha 10, Bandung
40132, Indonesia}
\address[icmp]{Indonesia Center for Theoretical and Mathematical Physics
(ICTMP), Department of Physics, Institut Teknologi Bandung Jl. Ganesha 10,
Bandung 40132, Indonesia}
\address[ipb]{Theoretical Physics Division, Department of Physics, Bogor
Agricultural University, Kampus IPB Darmaga, Bogor,16680 Indonesia}
\address[gftk]{Group for Theoretical and Computational Physics,
Research Center for Physics, Indonesian Institute of Sciences,
Kompleks Puspiptek Serpong, Tangerang 15310, Indonesia}
\address[ui]{Department of Physics, University of Indonesia,
Kampus UI Depok, Depok 16424, Indonesia\\}

\begin{abstract}
The impact of damping effect and external forces to the DNA breathing is
investigated within the Peyrard-Bishop model. In in the continuum limit, the
dynamics of the breathing of DNA is described by the forced-damped nonlinear
Schrodinger equation and studied by means of variational method.
The analytical solutions are obtained for special cases. It is shown that the
breather propagation is decelerated in the presence of damping factor without
the external force, while the envelope velocity and the amplitude
increase significantly with the presence of external force. It is particularly
found that the higher harmonic terms are enhanced when the periodic force is
applied. It is finally argued that the external force accelerates the DNA
breathing.
\end{abstract}


\end{frontmatter}

\section{Introduction}
\label{intro}

It is well known that the biological molecule functions such as
transcription and replication can not be explained only by their
static structure but also by their dynamical behavior
\cite{peyrard}. The transcription processes begun with separation
 of the double helix into a single helix  called
denaturation process \cite{yakushevich}. The famous model for
explaining thermal denaturation is the Peyrard-Bishop (PB) model
\cite{pbd1}. The model consists of two chains connected by Morse
potential representing the hydrogen (H) bonds. It has been shown
that the H bonds stretching depends on the coupling constant and
temperature, while in the continuum  limit satisfies the Nonlinear
Schrodinger equation. The solution of Nonlinear Schrodinger
equation is implies for a self-focusing case could  initiate the
denaturation \cite{pbd1}. The intensive studies on PB model have
been done for example, the molecular dynamics calculation
\cite{pbd2}, including anharmonics nearest-neighbor stacking
interaction \cite{pbd3}, connection with conformation of local
denaturation\cite{sung}, the effects of stacking interactions
\cite{kim} and the Monte-Carlo simulation \cite{ares}.

Even under physiological conditions, the DNA double-helix
spontaneously denatures locally, opening up and fluctuating, and
have a large amplitude localized excitations called DNA breathing
\cite{metzler,peyrard2}. It is well known that the thermal
denaturation of double strands DNA depend on the solution that
surrounding the DNA molecules \cite{voet}. Using this fact, it is
necessary to take into consideration that the solving water does
act as a viscous medium that could damp out DNA breathing. The
impact of viscosity was investigated by Zdrakovic et al
\cite{zts1}. The behavior of DNA dynamics in viscous solution is
described by the damped nonlinear Schrodinger equation. In the
other hand, the experiment showed that the double stranded DNA can
be separated by applying a fixed external force \cite{danilowicz}.
It was shown that the observed phase diagram for the unzipping of
double stranded DNA is much richer than the earlier suggestion
theoretical work. Therefore, it is necessary to improve the PB
model to take into account a viscosity and the external forced
simultaneously. To our knowledge, this problem has never been
reported elsewhere.

This paper discusses the viscous dissipation effect as well as an
external force acting to DNA in the Peyrard-Bishop models. In the
section ({\ref{sec:model}}) formulation of the impact of a damping
and external force is derived. The continuum approximation to
describe DNA breathing is described in section (\ref{sec:ca}).
Soliton solution based on the variational method is given in
section (\ref{sec:lagrange}). The breathing dynamics in term of
solitary waves is discussed in section ({\ref{sec:dbd}}). The
paper is ended with a summary.

\section{Peyrard-Bishop Model with damping and external forces}
\label{sec:model}
Following the PB model, the motion of DNA
molecules is represented by two degree of freedom $u_n$ and $v_n$
which correspond to the displacement of the base pair from their
equilibrium position along the direction of the hydrogen bonds
connecting the two base in pair in two different strands
\cite{pbd1}. Making a transformation to the center of mass
coordinate representing the in phase and out of phase transverse
motions, $X_n=(u_n+v_n)/\sqrt{2}$ and $Y_n=(u_n-v_n)/\sqrt{2}$
respectively, the Hamiltonian of the PB model is given by
\cite{pbd1}, \bea
 H & = & \sum_n \frac{1}{2M}(p_n)^2
 + \frac{\kappa}{2}(X_{n+1}-X_{n})^2  + \sum_n \frac{1}{2M}(P_n)^2
 \nonumber\\
 &&+ \frac{\kappa}{2}(Y_{n+1}-Y_{n})^2
 +\frac{D}{2}(e^{-\frac{\alpha}{2}Y_n}-1)^2 \; ,
 \label{eq:pb}
\eea
 where $D$ and $\alpha$ are the depth and inverse width of the potential respectively.
 The momentum $p_n=M \dot{X_n}$, $P_n=M\dot{Y_n}$ and $\kappa$ is the spring constant.

As mentioned above, the studies of PB models that included the
viscosity was done by adding the term $-\epsilon \gamma \dot{Y}_n$
in the equation of motion (EOM) \cite{zts1}. In ref. \cite{zts1} the
nonlinear Schrodinger equation with viscous effect was solved to
study the dynamics of DNA breathing. The interaction between the
system with it's environment lead dissipation of energy. This
means that the system is not longer conservative and reversible.
The corresponding Hamiltonian formulation for dissipative system
is called Caldirola-Kanai Hamiltonian in the form of a
time-dependent Hamiltonian which defined as follow \cite{um},
\be
  H= e^{-\gamma t} \frac{p^2}{2m} + e^{\gamma t}V(x)\; ,\
  \label{eq:kanai}
\ee where $\gamma =\eta/M$, $\eta$ is damping coefficient. The
model has been used to study the quantum dissipation such as the
quantization of an electromagnetic field inside a resonator filled
by dielectric medium \cite{dodonov}, study of susceptibility for
identical atoms subjected to an external force \cite{yeon},
coherent states for the damped harmonic oscillator\cite{yeon},
dissipative tunneling of the inverted Caldirola-Kanai oscillator
\cite{baskoutas} and functional integral for non-Lagrangian
systems \cite{kochan}.

In principle one can extend their approach to describe the
denaturation processes in a dissipation system. We propose that
the PB model with the damping effect and an external driving force
$F(t)$ is defined as follow, \bea
    H_x &=& \sum_n  e^{-\gamma t} \frac{p_n^2}{2m}
 + e^{\gamma t}\frac{\kappa}{2}(X_{n+1}-X_{n})^2 + e^{\gamma t} F_n(t)X_n  \; , \\
   H_y & = & \sum_n  \frac{e^{-\gamma t}}{2M} (P_n)^2 + \frac{e^{\gamma t}\kappa}{2} (Y_{n+1}-Y_{n})^2
 \nonumber\\  &&
 +\frac{e^{\gamma t}D}{2} (e^{-\frac{\alpha}{2}Y_n}-1)^2 +  e^{\gamma t} F_n(t)Y_n \; ,
 \label{eq:pbdF}
\eea Here  $F_n(t)$ is a conservative force. Substituting into the
Hamilton equation yield, \bea
 \dot{Y}_n&=&\frac{\pd H_y}{\pd P_n}= e^{-\gamma t} \frac{P_n}{M}
 \nonumber\\
 \dot{P}_n&=&-\frac{\pd H_y}{\pd Y_n}= e^{\gamma t}\kappa (Y_{n+1}-2Y_n+Y_{n-1}) \nonumber\\
        &+&  e^{\gamma t}\frac{\alpha D}{2}(e^{-\frac{\alpha}{2}Y_n}-1) + e^{-\frac{\alpha}{2}Y_n}
        + e^{\gamma t} F_n(t) \; .\
 \label{eq:eom1}
\eea Next, by substituting first equation into the second equation
we find the corresponding EOM as follow, \bea
 M \ddot{Y}_n &+& M\gamma \dot{Y}_n =\kappa (Y_{n+1}-2Y_n+Y_{n-1})  \nonumber\\
        &+& \frac{\alpha D}{2}(e^{-\frac{\alpha}{2}Y_n}-1)e^{-\frac{\alpha}{2}Y_n}
        + F_n(t)
        \; .\
        \label{eq:eom2}
 \eea
In the mean time, the EOM for $X_n$ satisfies the
following damped harmonic oscillator with external forcing,
\be
 M \ddot{X}_n + M \gamma \dot{X}_n -\kappa (X_{n+1}-2X_n+X_{n-1})+ F_n(t)=0
        \; .\
        \label{eq:eom3}
 \ee
The damping term $m\gamma \dot{Y}_n$ is similar with \cite{zts1},
where they add a new force $F_{n}=-\gamma \dot{Y}_n$ in the EOM.

\section{Continuum Limit Approximation}
\label{sec:ca} The dynamical behavior of DNA breathing can be
studied by applying the continuum approximation on the equation
(\ref{eq:eom3}). We assume that the amplitude of oscillation is
small and the nucleotide oscillate around the bottom of the Morse
potential but large enough due to nonlinear effect. We use the
following approximation \cite{peyrard,zts1,zts2},
\be
   Y_n = \epsilon \Psi_n ; (\epsilon \ll 1) \; .\
   \label{eq:aprok}
\ee Substituting Eq.(\ref{eq:aprok}) into Eq.(\ref{eq:eom2}) and
retained up to the third order of the Morse potential, we get \bea
  \ddot{\Psi}_n &+& \gamma \dot{\Psi} = \omega_0^2
  (\Psi_{n+1}-2\Psi_n+\Psi_{n-1}) \nonumber\\ &+& C_m^2(\Psi_n
  + \epsilon a_1 \Phi_n^2 + \epsilon^2 a_2^2 \Psi_n^3) + \tilde{F}_n\; ,\
  \label{eq:gerak1}
\eea where $\omega_0^2=\frac{\kappa}{M}$, $C_m^2=\frac{\alpha^2
\bar{D}}{2M}$, $a_1=-\frac{3}{4}\alpha$ , $a_2=\alpha
\sqrt{\frac{7}{24}}$, $\bar{D}=1/N\sum_n^N D_n$ is the average
value of $D$ and $\tilde{F}_n=F_n(\epsilon \Psi_n )$. For a
relatively long DNA chain, this equation can be simplified by
taking full continuum limit approximation which should be valid as
long as the solution under consideration changes rather slowly and
smoothly along with DNA \cite{yakushevich}. This approximation
yields, \be
  \frac{\pd^2 \Psi}{\pd t^2} + \gamma \frac{\pd \Psi}{\pd t} =
  C_0^2
  \frac{\pd^2 \Psi}{\pd x^2} + C_m^2(\Psi
  + \epsilon a_1 \Phi^2 + \epsilon^2 a_2^2 \Psi^3) + \tilde{F}(x,t)\; ,\
  \label{eq:gerak}
\ee where $C_0^2=\omega_0^2 l^2$ and  $l$ is a length scale. The
term of  $ \epsilon \tilde{F}(x,t)  $ and $\epsilon \gamma \pd
\Psi/ \pd t $ can be treated as a perturbation by assuming that it
contributes small enough to the whole DNA motion which should be
dominated by the first and second terms yielding a solution of
$\Psi \sim \e^{i ( k x - \omega t)}$. Then, using the multiple
scale expansion method, namely by expanding the associated
equation into different scale and time spaces \cite{whitham}, \bea
    \Psi     & = & {\Psi}^{(0)} + \epsilon {\Psi}^{(1)} + \cdots \;
    ,\label{eq:multipel1} \\
     \frac{\pd}{\pd t}&=& \frac{\pd}{\pd t_0} + \epsilon \frac{\pd}{\pd t_1} + \cdots \;
     ,\label{eq:multipel2} \\
    \frac{\pd}{\pd x}&=& \frac{\pd}{\pd x_0} + \epsilon \frac{\pd}{\pd x_1} + \cdots \; .
    \label{eq:multiple3}
\eea and by substituting this expansion into Eq. (\ref{eq:gerak}),
one obtains, \bea
  0 && =  \epsilon^0\left[ \frac{\pd^2 \Psi^{(0)}}{\pd t_0^2}- C_0^2 \frac{\pd^2 \Psi^{(0)}}{\pd x_0^2} - C_m^2 \Psi^{(0)}\right] \nonumber \\
  && + \epsilon^1 \left[ \frac{\pd^2 \Psi^{(1)}}{\pd t_0^2} + 2 \frac{\pd^2 \Psi^{(0)}}{\pd t_0 \pd t_1}
    -C_0^2 \frac{\pd^2 \Psi^{(1)}}{\pd x_0^2} + \gamma \frac{\pd \Psi^{(1)}}{\pd t_0} \right.  \nonumber \\
  && \; \; \; \; \; \; \; \;  \left.
     - 2 C_0^2 \frac{\pd^2 \Psi^{(0)}}{\pd x_0 \pd x_1}
     + C_m^2 \Psi^{(1)}-\frac{3}{2} C_m^2 (\Psi^{(0)})^2 +  \tilde{F} \right] \nonumber\\
  && + \epsilon^2 \left[ \frac{\pd^2 \Psi^{(0)}}{\pd t_1^2} + 2 \frac{\pd^2 \Psi^{(1)}}{\pd t_0 \pd t_1}
    -C_0^2 \frac{\pd^2 \Psi^{(0)}}{\pd x_1^2} \right.  \nonumber \\
  && \; \; \; \; \; \; \; \;  \left.
    -2 C_0^2 \frac{\pd^2 \Psi^{(1)}}{\pd x_0 \pd x_1}
    + \frac{7}{6} C_m^2 (\Psi^{(0)})^3  +2 C_m^2 \Psi^{(0)}\Psi^{(1)} \right] + ...
    \;.
 \label{eq:ekxpansi}
\eea Since the first and second terms in the leading order (LO)
and next to leading order (NLO) terms of Eq.(\ref{eq:ekxpansi})
provide the harmonic solutions, while the remaining terms lead to
non-harmonic solutions, then it is reasonable to consider, \bea
    \Psi^{(0)}(x_1,t_1)= \Psi^{(1)}(x_1,t_1) \, \e^{i(kx_0-\omega_0 t_0)} +cc \; ,
    \label{eq:xp0} \\
    \Psi^{(1)}(x_1,t_1)= \Psi^{(0)}(x_1,t_1) + \Psi^{(2)}(x_1,t_1) \, \e^{2i(kx_0-\omega_0 t_0)} +cc\; ,
    \label{eq:xp1} \\
    \tilde{F}=\tilde{f}(x_1) \, \e^{i(kx_0-\omega_0 t_0)} \; .
    \label{eq:fexp}
\eea Note that the time dependence on the right hand side of Eq.
(\ref{eq:fexp}) is introduced temporarily just for the sake of
convenience, and there are additional charge conjugate terms in
each equation. These lead to the order by order EOM, 
\bea
    && {\Psi}^{(0)} = 3  \left| {\Psi}^{(1)} \right|^2  \; ,
     \label{eq:fnls0} \\
   && i\frac{\pd {\Psi}^{(1)}}{\pd \tau} +  \Lambda_1\frac{\pd^2 {\Psi}^{(1)}}{\pd \xi^2}
   + \Lambda_2 \frac{\pd \Psi^{1}}{\pd \tau}
    \; \; \; \;  + \Lambda_3  \left| {\Psi}^{(1)} \right|^2 {\Psi}^{(1)}= \tilde{f} \;
    ,
    \label{eq:fnls1}\\
    && {\Psi}^{(2)} = \frac{1}{2} \left| {\Psi}^{(1)} \right|^2  \; ,
    \label{eq:fnls2}
\eea where $\Lambda_1=C_0 \bar{D}/(2 C_m^3)$, $\Lambda_2=\gamma/(2
C_m) $, $\Lambda_3=2 \bar{D}/C_m $,$\tau \equiv t_1=\epsilon t_0$
, $\xi \equiv x_1 - ({C_0 k}/{C_m}) t_0$, $x_1=\epsilon x_0$ and a
dispersion relation $\omega_0^2=\bar{D}+C_0^2k^2$. The EOM of
${\Psi}^{(1)}$ is nothing else than the forced-damped nonlinear
Schrodinger (FDNLS) equation.

\section{Variational Methods}
\label{sec:lagrange}

In this section we discuss the solution of the FDNLS by means of
variational methods  based on the Lagrangian formulation. It is
well know that for the case with $\gamma=0$ and $\tilde{f}=0$, the
Nonlinear Schrodinger equation admits the following traveling wave
solution \cite{yakushevich,zts1,zts2},
\be
  {\Psi}^{(1)}(\xi,\tau) = A_0 \, \mathrm{sech} \left[ \frac{1}{L} \, (\xi-u_e \tau ) \right]
  \e^{-i(\tilde{k}\xi-\tilde{\omega}\tau)}  \; ,
  \label{eq:solhomogen}
\ee where,
\bea
   &&A_0=\sqrt{\frac{[( u_e^2 - 2 u_e u_c )]}{(2 \Lambda_1\Lambda_3 )}} \\
  && L =\frac{\sqrt{2 \Lambda_1}}{\sqrt{u_e^2-2u_e u_c}}\\
  && \tilde{k} = \frac{u_e}{2 \Lambda_1} \\
 && \tilde{\omega} =\frac{u_e u_c}{2\Lambda_1}
 \label{eq:koefisien}
\eea Here, $u_e$ is the envelope wave velocity and $u_c$ is the
carrier wave velocity, satisfying $ u_e^2 - 2 u_e u_c
>0$.  By using Eq.(\ref{eq:multipel1}), Eq.(\ref{eq:xp0}),
Eq.(\ref{eq:xp1})  we obtain the soliton solution as follow,
\be
  \Psi(x,t)=2\Psi^{(1)}\cos(kx-\omega_0t) + \epsilon |\Psi^{(1)}|^2 \left[ 3+ \cos(2(kx-\omega_0t))\right] \; ,
  \label{eq:gelembung}
\ee where $\Psi^{(1)}$ is described by (\ref{eq:solhomogen}).

 Based
on the corresponding variational methods to solve the
damped-forced nonlinear Schroedinger equation, one can use the
solution (\ref{eq:solhomogen}) as the related basic form and
considering its amplitude, width, phase velocity and the position
of the soliton  to be time dependent \cite{whitham, grimshaw,
mertens}. For convenient, let us write the 1-soliton in the
following form,

\be
   \Psi^{(1)}(\xi,\tau)=\eta(\tau) \mathrm{sech}[\eta(\xi+\zeta(\tau))] \exp
   \left(-i[\theta(\tau)\xi+\phi(\tau)]\right)
   \; .\
   \label{eq:single1}
\ee

The dynamics of $\eta$, $\theta$, $\zeta$ and $\phi$ function can
be obtained by using the variational methods. The Lagrangian that
satisfy the Euler-Lagrange equation \cite{whitham},
\be
  \frac{\pd}{\pd \tau}\left(\frac{\pd \l}{\pd
  \Psi^{(1)*}_\tau} \right)
  +\frac{\pd}{\pd \xi}\left (\frac{\pd \l}{\pd \Psi^{(1)*}_\xi}
  \right)
  - \frac{\pd \l}{\pd \Psi^{(1)*}}= 0 \; .\
  \label{eq:euler}
\ee for the FDNLS is given by, \bea
  \l &=&\frac{i}{2}(\Psi^{(1)}_\tau \Psi^{(1)*}-\Psi^{(1)*}_\tau
  \Psi^{(1)})-\Lambda_1 \mid \Psi^{(1)}_\xi \mid^2 + \Lambda_3
  \mid \Psi^{(1)} \mid^4 \nonumber\\
  &+& \frac{\Lambda_2}{2}(\Psi^{(1)}_\tau \Psi^{(1)*}-\Psi^{(1)*}_\tau
  \Psi^{(1)})-(\tilde{f}\Psi^{(1)*}+\tilde{f}^* \Psi^{(1)}) .\
  \label{eq:lagrange1}
\eea Substituting equation (\ref{eq:single1}) into the equation
(\ref{eq:lagrange1}) and using $L=\int_{-\infty}^{\infty} \l
d\xi$, $\int_{-\infty}^{\infty} \mathrm{sech}(a\xi) d\xi=\pi/a$
and $\int_{-\infty}^{\infty} \mathrm{sech}^2(a\xi)\tanh(a\xi)
d\xi=0$ yield, \be
  L = 2\eta \zeta \dot{\theta} + \eta \dot{\phi} + 2i\Lambda_2
  \eta \zeta \dot{\theta} + i \Lambda_2 \eta \dot{\phi} + 4\Lambda_1 \eta \theta^2 + \frac{4}{3}\Lambda_3
  \eta^3 - \eta \bar{F} \; ,
  \label{eq:lagrange2}
\ee where
\be
   \bar{F}= \int_{-\infty}^{\infty}
   \left(\tilde{f} \e^{i[\theta(\tau)\xi+\phi(\tau)]}+\tilde{f}^*
    \e^{-i[\theta(\tau)\xi+\phi(\tau)]} \right)\mathrm{sech}[\eta(\xi-\zeta(\tau)] d\xi .\
   \label{eq:gayaluar}
\ee Eq. (\ref{eq:lagrange2}) is the Lagrange function in term of
$\theta$, $\eta$, $\phi$ and $\zeta$. The EOM can
be easily obtained by solving the Euler-Lagrange equation,
\be
  \frac{d}{dt}\left(\frac{\pd L}{\pd \dot{X}}\right) -\frac{\pd
  L}{\pd X}=0 \; ,
  \label{eq:euler2}
\ee where $X=(\eta, \theta, \phi, \zeta)$. Substituting the
Lagrange function given by Eq.(\ref{eq:lagrange2}) into
Eq.(\ref{eq:euler2}) leads to the following, \bea
    2(1+i\Lambda_2)\eta \dot{\zeta}+2(1+i\Lambda_2)\zeta
    \dot{\eta} &=& 8\Lambda_1 \eta \theta -\eta \frac{\pd \bar{F}}{\pd
    \theta}  \label{eq:EOM1a}    \\
     2(1+i\Lambda_2)\zeta \dot{\theta} +
     (1+i\Lambda_2)\dot{\phi} &=&4(\Lambda_1 \theta^2 - \Lambda_3
     \eta^2)- \eta \frac{\pd \bar{F}}{\pd \eta}-\bar{F}  \label{eq:EOM1b} \\
      (1+i\Lambda_2)\dot{\eta} &=& -\eta \frac{\pd \bar{F}}{\pd \phi}  \label{eq:EOM1c} \\
       2(1+i\Lambda_2)  \dot{\theta}&=& - \frac{\pd \bar{F}}{\pd \zeta} \,
  \label{eq:EOM1d}
\eea Further, by substituting Eq.(\ref{eq:EOM1c}) and
Eq.(\ref{eq:EOM1d}) into Eq.(\ref{eq:EOM1a}) and
Eq.(\ref{eq:EOM1b}) respectively, we find
\bea
    (1+i\Lambda_2)\eta \dot{\zeta}-4\Lambda_1 \theta &=& 2 \eta \frac{\pd \bar{F}}{\pd
    \eta} - \frac{\pd \bar{F}}{\pd \theta}  \label{eq:ODE1a}
    \\
     (1+i\Lambda_2)\dot{\phi} -4(\Lambda_1 \theta^2 - \Lambda_3\eta^2)
     &=&  \zeta \frac{\pd \bar{F}}{\pd \zeta}-\eta \frac{\pd \bar{F}}{\pd \eta}-\bar{F} \; .
     \label{eq:ODE1b}
\eea Generally, the analytic solution of the equation
(\ref{eq:EOM1a})-(\ref{eq:EOM1d}) can not be obtained, but it is
still possible to find it for special condition. In the next
section we restrict ourselves to special case of $\tilde{f}$.

First, let us consider very special case in which the damping
factor and external force are vanish ($\Lambda_2=0$ and
$\bar{F}=0$). As a result , Eq.(\ref{eq:EOM1c}) show that the
function of $\eta=\eta_0$ is a constant and Eq.(\ref{eq:EOM1d})
yield $\theta=\theta_0$ which is also a constant.
Eq.(\ref{eq:EOM1a}) and Eq.(\ref{eq:EOM1b}) lead to a simple
solutions, \bea
  \zeta &=& 4 \Lambda_1 \theta_0 \tau \; \\
  \phi &=& 4 (\Lambda_1 \theta_0^2-\Lambda_3 \eta_0^2)\tau \; .
  \label{eq:SolnodampF}
\eea Such that single soliton have the form of,
\be
   \Psi^{(1)}(\xi,\tau)=\eta_0 \mathrm{sech}\left[\eta_0(\xi+4 \Lambda_1 \theta_0 \tau)\right]
   \exp \left(-i[\theta_0 \xi+4 (\Lambda_1 \theta_0^2-\Lambda_3
   \eta_0^2)\tau]\right)
   \; \, .\
   \label{eq:single2}
\ee Clearly, this is a single soliton solution of the conventional
Nonlinear Schrodinger equation.

In the second case, let us ignore the external forces
($\bar{F}=0$). Again, as a result, Eq.(\ref{eq:EOM1c}) show that
the function of $\eta=\eta_0$ is a constant and
Eq.(\ref{eq:EOM1d}) yield $\theta=\theta_0$, constant as well. The
time dependent variables are $\zeta$ that represent of the
velocity of soliton and $\phi$ the phase of soliton. The equation
of motion have a simple form as follow, \bea
    (1+i\Lambda_2)\frac{d \zeta}{d \tau}- 4\Lambda_1 \theta_0 &=&0
    \label{eq:ODE2a} \\
    (1+i\Lambda_2)\frac{d \phi}{d \tau}- 4\left(\Lambda_1
    \theta_0^2 - \Lambda_3 \eta_0^2 \right) &=&0
    \label{eq:ODE2b}
\eea This yield,
\bea
    \zeta(\tau) &=&4\frac{\Lambda_1
    \theta_0}{(1+\Lambda_2^2)}\left(1-i\Lambda_2\right)\tau
    \label{eq:SolODE2a} \\
    \phi(\tau) &=&  4\frac{\left(\Lambda_1 \theta_0^2 - \Lambda_3 \eta_0^2
    \right)}{(1+\Lambda_2^2)}\left(1-i\Lambda_2\right)\tau
    \label{eq:SolODE2b}
\eea Finally, the single soliton solution is,

\bea
   \Psi^{(1)}(\xi,\tau) &=&\eta_0 \mathrm{sech}[\eta_0(\xi+\bar{\zeta}\tau)+ i\eta_0 \bar{\zeta}
   \Lambda_2 \tau]\nonumber\\
   &\times& \exp\left(-\bar{\phi}\Lambda_2\tau \right)\exp\left(-i(\theta_0\xi+\bar{\phi}\tau)\right)\;  ,
   \label{eq:single3}
\eea where $\bar{\zeta}=4\Lambda_1\theta_0/(1+\Lambda_2^2)$ and
$\bar{\phi}=4(\Lambda_1\theta_0^2-\Lambda_3\eta_0^2)/(1+\Lambda_2^2)$.

For the third case, we consider another simple case namely when
$\bar{F}=\bar{F}_0$ is a constant. Again we get $\eta=\eta_0$  and
$\theta=\theta_0$ are constants. The EOMs in Eq.(\ref{eq:ODE1a}) and
Eq.(\ref{eq:ODE1b}) have simple forms as follow, 
\bea
    (1+i\Lambda_2)\frac{d \zeta}{d \tau}- 4\Lambda_1 \theta_0 &=&0
    \label{eq:ODE3a} \\
    (1+i\Lambda_2)\frac{d \phi}{d \tau}- 4\left(\Lambda_1
    \theta_0^2 - \Lambda_3 \eta_0^2 \right) &=& -\bar{F}_0
    \label{eq:ODE3b}
\eea This yields, \bea
    \zeta(\tau) &=&4\frac{\Lambda_1
    \theta_0}{(1+\Lambda_2^2)}\left(1-i\Lambda_2\right)\tau
    \label{eq:SolODE3a} \\
    \phi(\tau) &=&  \left[\frac{4\left(\Lambda_1 \theta_0^2 - \Lambda_3 \eta_0^2
    \right)}{(1+\Lambda_2^2)}-\frac{\bar{F}_0}{(1+\Lambda_2^2)} \right]\left(1-i\Lambda_2\right)
    \tau
    \label{eq:SolODE3b}
\eea Finally, the single soliton solution for this case is, \bea
   \Psi^{(1)}(\xi,\tau) &=&\eta_0 \exp\left(-\bar{\phi}\Lambda_2\tau \right) \exp\left(F^*\tau\right) \nonumber\\
   &\times&  \mathrm{sech}[\eta_0(\xi+\bar{\zeta}\tau)+ i\eta_0 \bar{\zeta}
   \Lambda_2 \tau] \exp\left[-i(\theta_0\xi+\bar{\phi}\tau-F^*\tau)\right]\;  ,
   \label{eq:single3}
\eea

where $F^*=\bar{F}_0/(1+\Lambda_2^2)$. The external force express
in the positive exponential term, this show that it will increase
the amplitude of the soliton.

In the fourth case, we assume that the amplitude and the phase is
a constant ($\eta =\eta_0$) and ($\theta=\theta_0$), then the
EOM becomes, \bea
    (1+i\Lambda_2)\eta_0 \dot{\zeta}-4\Lambda_1 \theta_0 &=& 0 \label{eq:ODE4a}
    \\
     (1+i\Lambda_2)\dot{\phi} -4(\Lambda_1 \theta_0^2 - \Lambda_3\eta_0^2)
     &=&  \zeta \frac{\pd \bar{F}}{\pd \zeta}-\bar{F} \; .
     \label{eq:ODE4b}
\eea Further, let us assume that the external force is expressed
specifically by
\be
  \tilde{f}=f_0 \exp(-i\theta_0 \xi) \; ,
  \label{eq:external1}
\ee with $f_0$ is a constant. Substituting into
Eq.(\ref{eq:gayaluar}) yields, \bea
   \bar{F} &=& \int_{-\infty}^{\infty}
   \left(f_0 \e^{-i\theta_0 \xi} \e^{i[\theta_0 \xi+\phi(\tau)]}+ f_0
   \e^{i\theta_0 \xi} \e^{-i[\theta_0\xi+\phi(\tau)]} \right)\mathrm{sech}[\eta(\xi-\zeta(\tau)] d\xi
   \; \nonumber\\
    &=& \frac{4 f_0}{\pi}\cos(\phi)
   \label{eq:external2}
\eea Such that we find, \bea
     \dot{\zeta} &=& \frac{4\Lambda_1 \theta_0}{(1+i\Lambda_2)\eta_0} \label{eq:ODE4aa}
    \\
    \dot{\phi}
     &=& -\frac{4 f_0}{\pi(1+i\Lambda_2)} \cos(\phi)+ \frac{4(\Lambda_1 \theta_0^2 - \Lambda_3\eta_0^2)}
     {(1+i\Lambda_2)}  \; .
     \label{eq:ODE4bb}
\eea The Eq.(\ref{eq:ODE4bb}) will be solved numerically. It is
important to note that the solution of Eq.(\ref{eq:ODE4bb}) is a
complex function i.e. $\phi=\phi_R + i \phi_I$, such that the
soliton profile is given by,
\be
   \Psi^{(1)}(\xi,\tau) =\eta_0 \exp\left(\phi_I \tau \right) \mathrm{sech}[\eta_0(\xi-\bar{\zeta}\tau)+ i\eta_0 \bar{\zeta}
   \Lambda_2 \tau] \exp\left[-i(\theta_0\xi+ \phi_R \tau)\right]\;  ,
   \label{eq:single3}
\ee

\section{Nonlinear Dynamics of DNA Breathing}
\label{sec:dbd}

The external effects of DNA usually give inhomogeneities in the
DNA model. The inhomogeneity in stacking energy is found to
modulate the width and speed of the soliton which depend on the
nature of the inhomogeneity \cite{deniel}. The author used the
dynamic plane-base rotator model by considering angular rotation
of bases in a plane normal to the helical axis. They found that
the DNA dynamics is governed by a perturbed sine-Gordon equation.
In this paper we found that the inhomogeneities of the DNA
breathing dynamics governed by the forced-damped Nonlinear
Schrodinger equation. The solutions of the homogeneous case
represent a large amplitude with localized oscillatory mode
appears as a good explanation of the breathing of DNA and must be
spontaneously formed \cite{peyrard,yakushevich,peyrard2}.

Let us discuss the previously considered four simple cases. In the
first case, it's shown  by using transformation $\eta_0=A_0$ and
$\theta_0=1/2u_e/\Lambda_1$ that the solution is just the
Eq.(\ref{eq:solhomogen}). We simulate the solution for the model
parameter given by $\kappa= 8 Nm$, $M=5.1\times 10^{-25} {kg} $,
$\alpha =2 \times 10^{10} m^{-1}$, $D=0.1 eV$ and $l=3.4 \times
10^{-10} m$ is length scale \cite{zts1}. The system of unit
($A^{o}$,$eV$) defines a time unit ($t.u.$) equal to $1.021\times
10^{-14}s$ \cite{tabi}. The solution demonstrate a sort of a
modulated solitonic wave where the hyperbolic and cosine terms 
correspond to the wave number of the envelope and the carrier
wave respectively.

\begin{figure}[t]
 \centering
 \includegraphics[width=1 \textwidth]{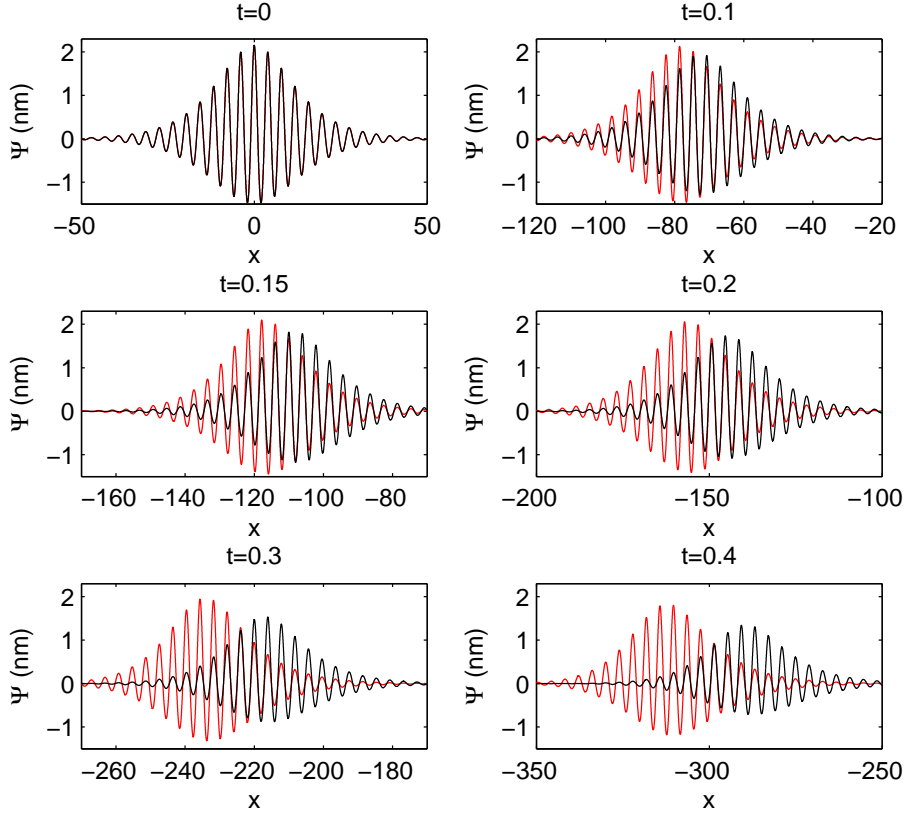}
 \caption{The DNA breathing in the second case where $x$ denotes the continuum
base pair in the present model (black) and the original PB model (red) with
$g=0.05$ and $F=0$.}
 \label{fig:casetwo}
\end{figure}

In second case with $\bar{F}=0$ and damping constant $\gamma=0.05
kg/s$, the behavior of the breathing is depicted in
Fig.({\ref{fig:casetwo}}). The figure show that the the breathing
propagation along the DNA molecule is effected by the damping. It
is shown from the figure that the damping term decelerates the
propagating soliton while retaining its amplitude profile. This
indicates that the corresponding damping term does not affects the
soliton mass.

\begin{figure}[t]
 \centering
 \includegraphics[width=1\textwidth]{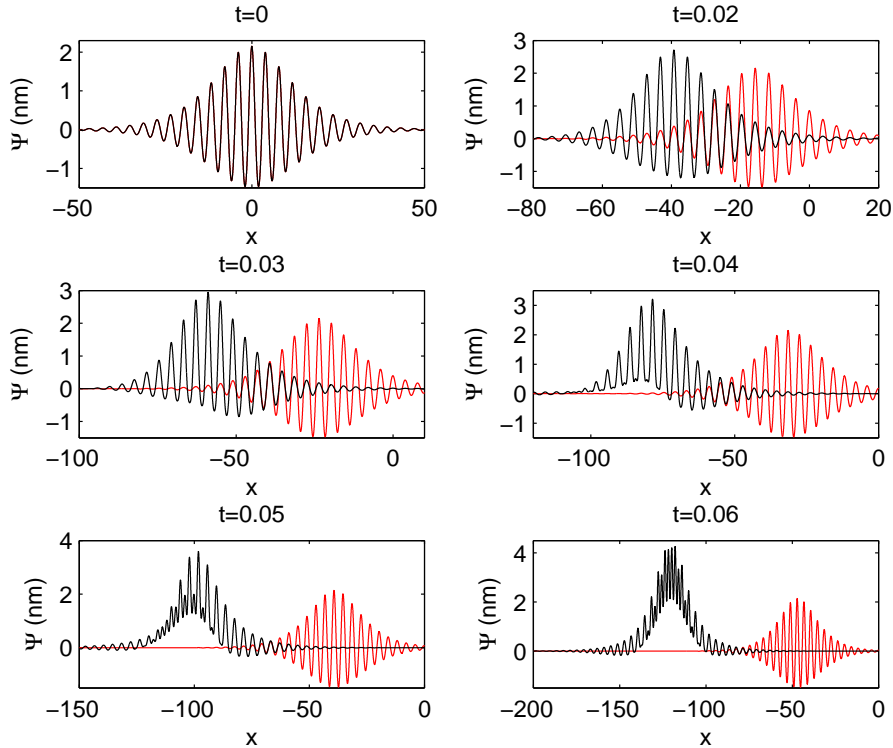}
 \caption{The solitonic solution of DNA breathing in the model with
damping effect and external force $\bar{F}$ (black) and in the original PB
model (red) with  $\gamma = 0.05$ and  $\bar{F}=15$.}
 \label{fig:casethree}
\end{figure}

Now, consider the solution with the present of the damping effect
and $\bar{F} \neq 0$ where the solution is depicted in
Fig.({\ref{fig:casethree}}). The figure is generated with
$\bar{F}=15 pN$ and a damping factor  $\gamma=0.05 kg/s$. The
external force tend to increase the breathing amplitude and damped
out by the damping effect. The consequences of this matter, the
present of external force tend to increase the envelope velocity
significantly. The velocity is increase around $\Delta u_e =\Delta
\xi/\Delta \tau \sim  1 order$ in the present of external force
$\bar{F}=15pN $.

Our result is supported by the report given in ref.
\cite{rouzina1} results. In a cell, DNA strands are separated by
the external force \cite{rouzina1, rouzina2}, or in chemical
terms, by enzymes whose interactions with DNA make strands
separation thermodynamically favorable at ambient temperature
\cite{sasaki}. They showed that the two strands of double-stranded
DNA can be separated (unzipped) by the application of $~15 pN$
force applied at room temperature. Their model predicts that the
melting temperature should be a decreasing function of applied
force. The paper show that the external force can increase the
amplitude of the breathing and may separate the double helix into
single helix.

\begin{figure}[t]
 \centering
 \includegraphics[width=.8\textwidth]{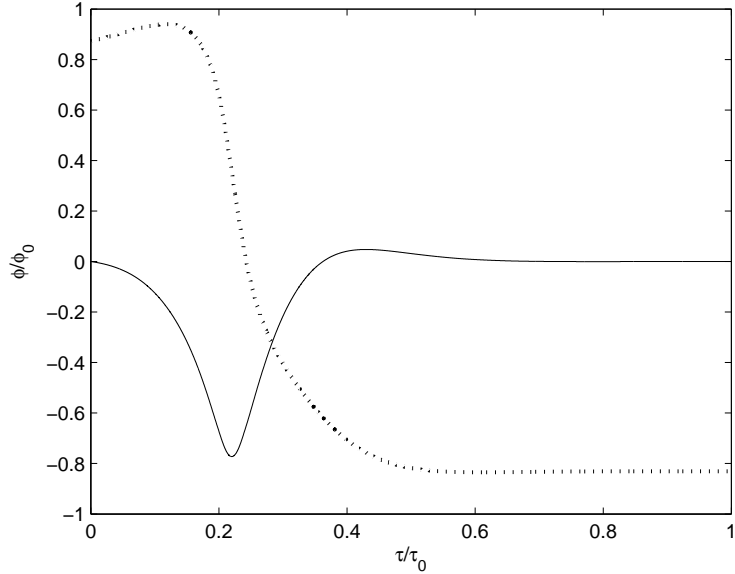}
 \caption{The solution of Eq. (\ref{eq:ODE4bb}) where the real and imaginary
parts are shown by the dot and solid lines.}
 \label{fig:phinya}
\end{figure}

Finally, let us consider the fourth case. The behavior of this
case described by on the numerical solution of
Eq.(\ref{eq:ODE4bb}) which gives a complex function of $\phi$. The
real part of the solution is associated with the carrier wave,
while the imaginary part with the envelope velocity respectively.
The profile of the real part and the imaginary part can be seen in
Fig.({\ref{fig:phinya}}). The real part profile seems to have  an
anti-sigmoid-like function. In the mean time, the profile of the
imaginary part have negative value, which means that the solution
propagates to the left direction.

\begin{figure}[t]
 \centering
 \includegraphics[width=1.1\textwidth]{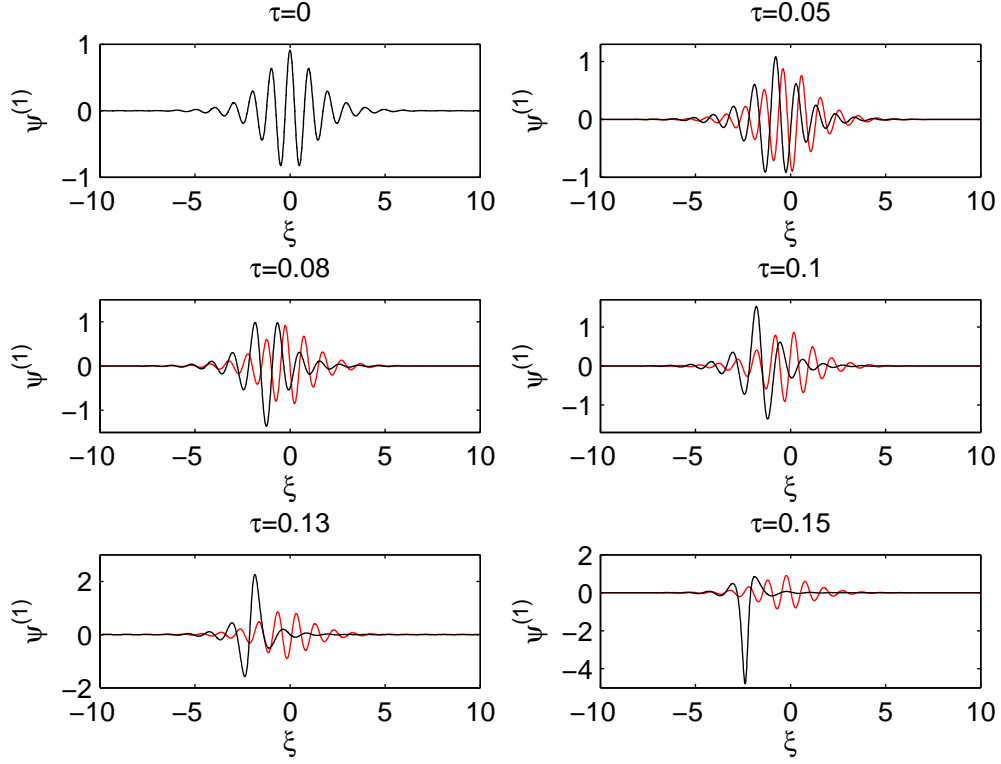}
 \caption{The solution of soliton in the FDNLS (black) and the NLS (red) with
$\gamma =0.05$, $\bar{F}=15\e^{i\xi}$, $v_e=10^{5}$ m/s and $u_c=4\times
10^{4}$ m/s.}
 \label{fig:psizda}
\end{figure}

At this point we can consider the solution of FDNLS equation by
using Eq.(\ref{eq:ODE4bb}) and this is depicted in
Fig.({\ref{fig:psizda}}). Soliton solution of FDNLS show an
increasing of the amplitude and its velocity generally tend to
increase with time. The amplitude is relatively high and the shape
of soliton tend to be step form during it's propagation.

\begin{figure}[t]
 \centering
 \includegraphics[width=1.1\textwidth]{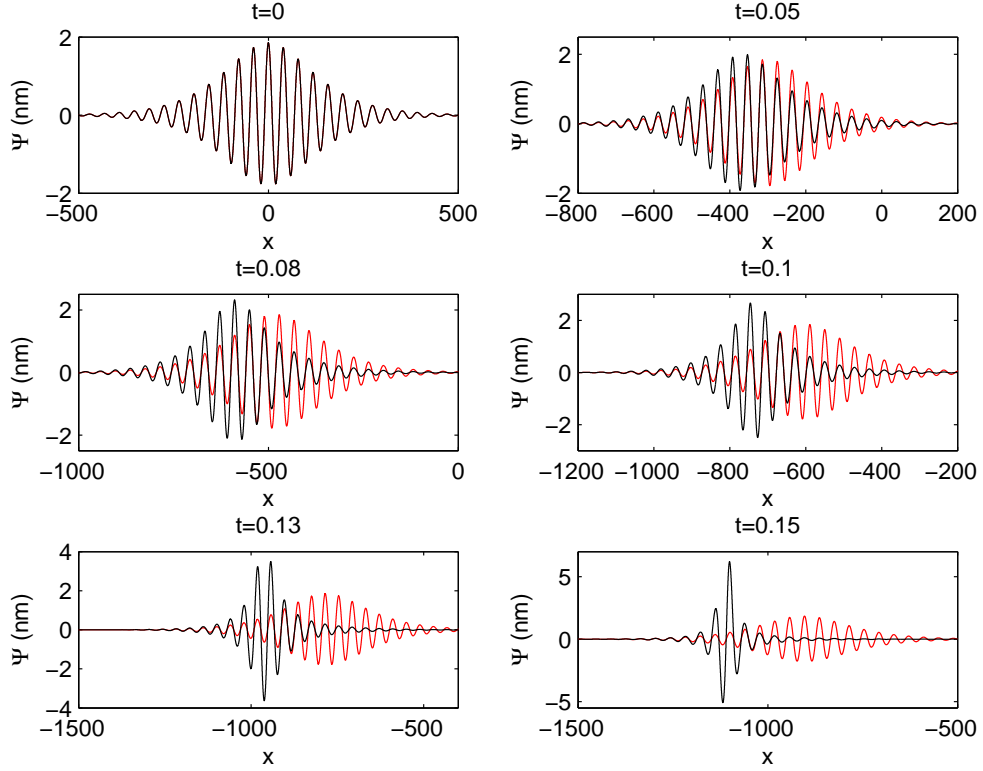}
 \caption{The DNA breathing propagation for the fourth case in the
present model (black) and the original PB model (red) with $\gamma
=0.05$, $\bar{F}=15\e^{i\xi}$, $v_e=10^{5}$ m/s and  $u_c=4\times 10^{4}$ m/s. }
 \label{fig:breathingzda}
\end{figure}

Now, the DNA breathing of the fourth case using the solution of
$\Psi^{(1)}$  given in Fig. ({\ref{fig:psizda}}) is depicted in
Fig.({\ref{fig:breathingzda}}). The amplitude of the DNA breathing
is around $2 nm$ where the result is similar with ref. \cite{zts1}
for the same parameters. The amplitude of the DNA breathing is
relatively increase  when its propagate to the left direction. The
high amplitude show that the DNA tend to be unzipped and finally
completely separate into single helix.

\begin{figure}[t]
 \centering
 \includegraphics[width=1.1\textwidth]{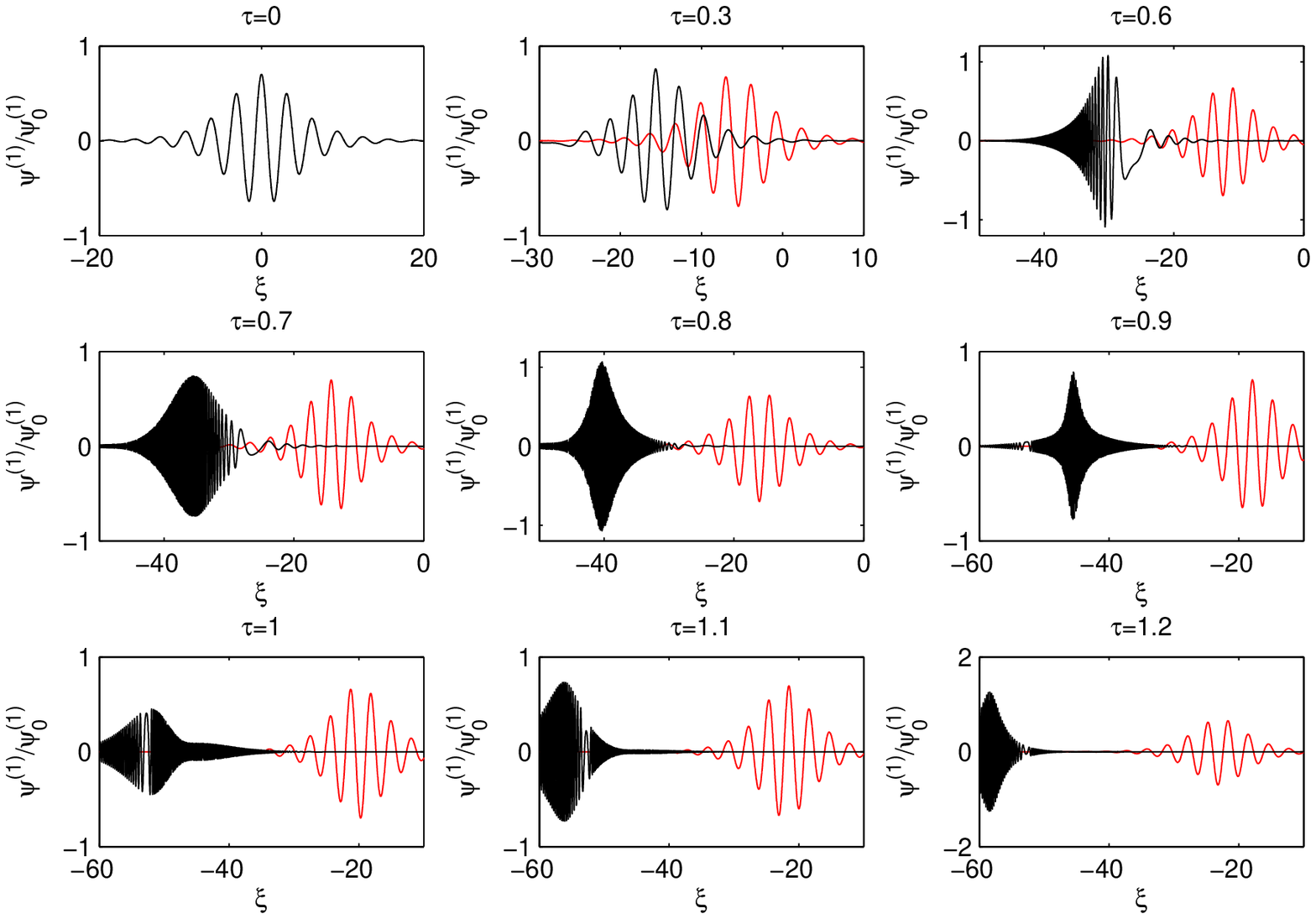}
 \caption{The solution of FDNLS (black) and NLS (red) with $\gamma
=0.05$, $\bar{F}=15\e^{i\xi}$, $K=4$ Nm, $v_e=10^{4}$ m/s and $u_c = 4 \times
10^{3}$ m/s.}
 \label{fig:psiman}
\end{figure}

Finally, it is interesting to note the case if we decrease the
envelope and carrier velocity respectively in one order. The
results is depicted in Fig.({\ref{fig:psiman}}). In this case, the
solution of FDNLS propagates more slower than the previous one. At
$\tau=0.3$ the soliton with periodic external force tend to
increase its amplitude and velocity. Further at $\tau=0.6$ the
amplitude increase significantly and the corresponding form is
change significantly. At $\tau=0.7$ the higher harmonic term is
developed and completely forms at $\tau=0.9$.  As the times goes
up the amplitude tend to decrease and disperse into wider form and
then the higher harmonic term is generated and increase the
amplitude. It is interesting to pointed out that the harmonic term
come from the the solution of Eq.(\ref{eq:ODE4bb}) that is a
nonlinear equation.

\begin{figure}[t]
 \centering
 \includegraphics[width=1.1\textwidth]{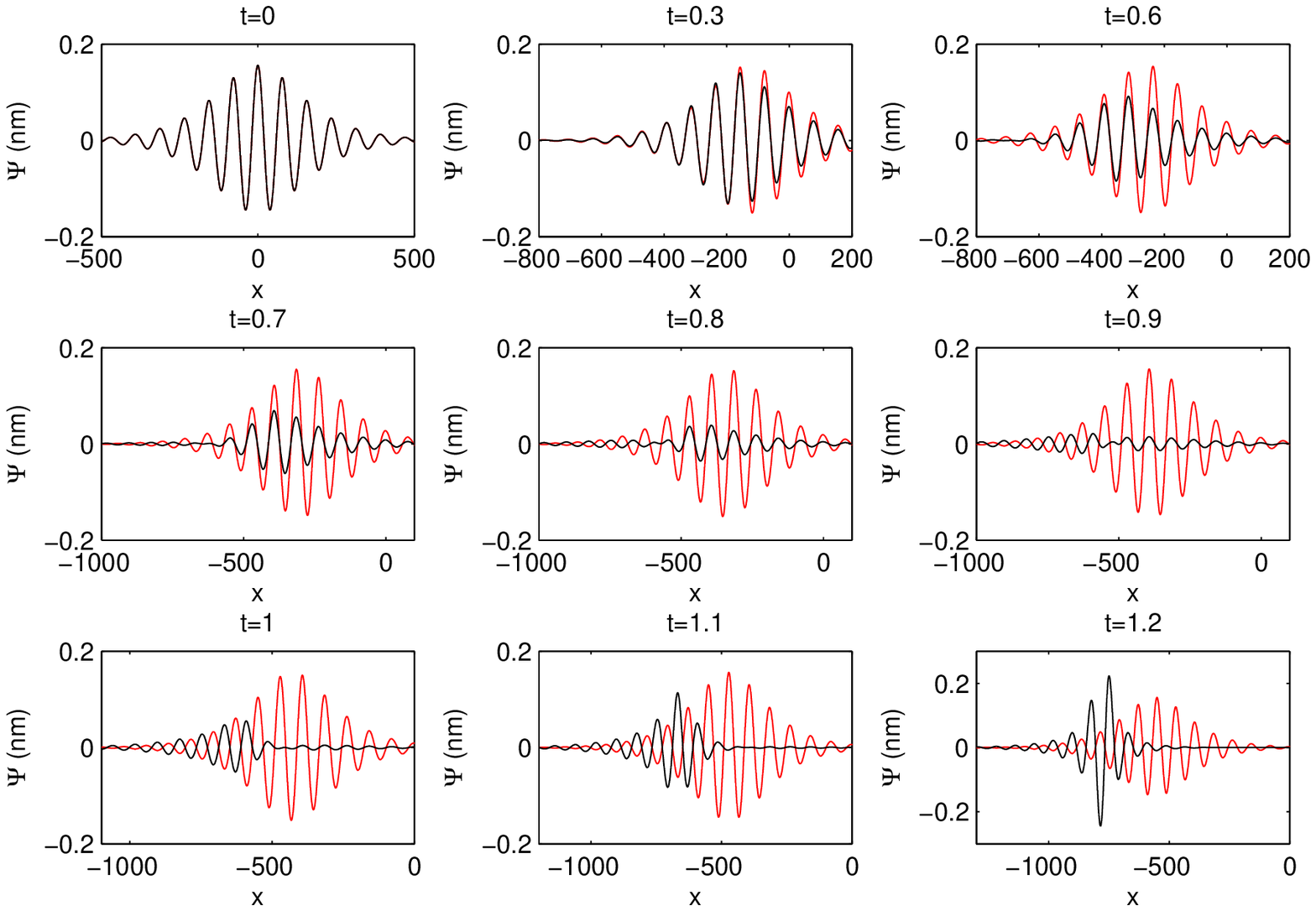}
 \caption{The DNA breathing in the model (black) and the original PB model
(red) with $\gamma =0.05$, $\bar{F}=15\e^{i\xi}$, $K=4$ Nm, $v_e=10^{4}$ m/s 
and $u_c=4\times 10^{3}$ m/s.}
 \label{fig:breathingman}
\end{figure}

The DNA breathing corresponds to the $\Psi^{(1)}$ above is depicted
in Fig.({\ref{fig:breathingman}}).  The result shows that the
amplitude is about $0.2 nm$ which is the same with the result
found in  ref. \cite{tabi} in which they used the
Forinash-Cretary-Peyrard model with helicosity taken into account.
They showed that the opening of the double helix of the DNA itself
is controlled by the resonance mode.

As the result shows that the periodic external force and the
damping effect generate a higher harmonic term in the dynamics of
FDNLS solution then it can be concluded that this phenomenon is
responsible for the dynamics of the breathing of the DNA. At this
point, we can see that in the early propagation process the DNA
breathing tend to decrease its amplitude and disperse into a wider
form. The condition changes when the higher harmonic term of FDNLS
soliton begin to develop which leads the amplitude of the DNA
breathing increases significantly.

\section{Summary}

The impact of viscous fluid and external forces to the Peyrard-Bishop DNA
breathing is investigated. We have proposed a PB model with the damping effect
and the external driving force which is described by an extended time-dependent
Caldirola-Kanai Hamiltonian. Taking full continuum approximation and using the
multiple scale expansion method, the EOM is nothing else than the FDNLS
equation. Assuming small perturbation of damping and external forces, the FDNLS 
equation can be solved using the variational methods. The analytical solution
have been obtained for only special cases. In the case with damping factor and
without external force, the breathing propagation is decelerated by the damping
effect. In the presence of external force, the velocity and the amplitude 
increase significantly. It is also found that the higher harmonic terms are
enhanced when the periodic force is applied.

These results shows that the external force contributes constructively
to accelerate the separation of double helix into single helix. More
comprehensive numerical investigation of the variational equation found in
this paper is still in progress.

\section*{Acknowledgments}
AS thanks the Group for Theoretical and Computational Physics LIPI
for warm hospitality during the work. This work is funded by the
Indonesia Ministry of Research and Technology and the Riset
Kompetitif LIPI in fiscal year 2011 under Contract no.
11.04/SK/KPPI/II/2011. FPZ thanks Research KK ITB 2011.

\bibliographystyle{elsarticle-num}
\bibliography{sulaiman}

\end{document}